\documentclass[referee]{raa}           
\usepackage{graphicx,times}
\usepackage{natbib}
\usepackage{amssymb,amsmath}
\usepackage{caption}
\usepackage[pagebackref=true]{hyperref}
\bibpunct{(}{)}{;}{a}{}{,}
\newcommand\M{\rm M}
\newcommand\A{\rm A}
\newcommand\B{\rm B}
\newcommand\HH{\rm H}
\newcommand\dd{\rm d}
\newcommand\p{\rm p}
\newcommand\fid{\rm fid}

\begin{document}
\title{Dynamical dark energy in light of cosmic distance measurements II: a study using current observations}

 \volnopage{ {\bf 20XX} Vol.\ {\bf X} No. {\bf XX}, 000--000}
   \setcounter{page}{1}

   \author{Xiaoma Wang
   \inst{1,2}, Gan Gu\inst{1,2}, Xiaoyong Mu\inst{1,2}, Shuo Yuan
      \inst{1},  Gong-Bo Zhao\inst{1,2,3}
   }

   \institute{National Astronomical Observatories, Chinese Academy of Sciences, Beijing, 100101, P.R.China, 
China; {\it gbzhao@nao.cas.cn}\\
        \and
             University of Chinese Academy of Sciences, Beijing, 100049, P.R.China\\
	\and
Institute for Frontiers in Astronomy and Astrophysics, Beijing Normal University, Beijing, 102206, P.R.China\\
\vs \no
   {\small Received 2024 Month Day; accepted 2024 Month Day}
}

\abstract{We extract key information of dark energy from current observations of BAO, OHD and $H_0$, and find hints of dynamical behaviour of dark energy. In particular, a dynamical dark energy model whose equation of state crosses $-1$ is favoured by observations. We also find that the Universe has started accelerating at a lower redshift than expected.
}

   \authorrunning{X. Wang et al. }           
   \titlerunning{Probing dynamical dark energy using current observations}  
   \maketitle

\section{INTRODUCTION} \label{sec:intro}

The physical origin of the cosmic acceleration remains unveiled since its discovery in $1998$ from observations of supernovae Type Ia (SNe Ia) \citep{1998AJ....116.1009R, 1999ApJ...517..565P}. Among all possible mechanisms, dark energy (DE) \citep{DE} and modified gravity (MG) \citep{MG} are two primary scenarios that have been extensively studied. Both DE and MG models can yield the same accelerating expansion of the Universe after the required tuning, and the expansion rate of the Universe is determined by the equation of state (EoS) $w$ of an effective dark fluid.

The EoS is generally a function of time, expressed by the scale factor $a$ or redshift $z$, and it is a ratio between the pressure $P$ and $\rho$ of the effective dark fluid. It is important to extract information of $w(z)$ from observations, because different DE or MG theory models can in principle be differentiated using the behaviour of $w(z)$. The $\Lambda$CDM model, in which dark energy is the vacuum energy, predicts $w=-1$. Although still allowed by observations, this model suffers from serious theory problems \citep{Weinberg:1988cp}, and has been challenged by the `Hubble tension' \citep{Hubble_tension}. In alternative models, $w$ generally evolves with time. For example, the single-scalar-field models of quintessence \citep{quintessence} and phantom \citep{phantom} predict $w>-1$ and $w<-1$ during the evolution, respectively, and in models with intrinsic degrees of freedom such as quintom \citep{quintom}, $w$ can cross the $-1$ boundary.

Observations of the cosmic distances, such as the baryonic acoustic oscillations (BAO) \citep{BAO98,BAO05a,BAO05b}, are less subject to systematics compared to probes of the cosmic structure formation, and the behaviour of $w(z)$ directly affects the cosmic distances, therefore the distance measurements are ideal observables for dark energy studies.

In this work, we learn the behaviour of $w(z)$ from current measurements of the cosmic distances based on the method developed in a companion paper \citep{GG24}, and find hints of interesting features of $w(z)$. We present the method and data used in this analysis in Sec. \ref{sec:method}, show the result in Sec. \ref{sec:result} and conclude in Sec. \ref{sec:conclusion}.

\section{Method and Data} \label{sec:method}

In this section, we briefly describe the method used for this work, as developed in a companion paper \citep{GG24}, and the datasets used for the analysis.

We start from parametrising the angular diameter distances\footnote{For current data, we find that keeping the expansion up to the $x^4$ term is a reasonable choice to balance between the bias and uncertainty of the reconstruction.} following \cite{Zhu:2014ica},
\begin{equation}
    \frac{D_{\A}(z)}{D_{\A,\fid}(z)}=\alpha_{0}\left(1+\alpha_{1}x+
    \frac{1}{2}\alpha_{2}x^{2}+
    \frac{1}{6}\alpha_{3}x^{3}+
    \frac{1}{24}\alpha_{4}x^{4}\right)\label{eq:para-DA},
\end{equation}
where $1+x\equiv D_{\A,\fid}(z)/D_{\A,\fid}(z_{\star})$, and subscript `fid' denotes the fiducial model, which is chosen to be a flat $\Lambda$CDM model with $\Omega_{\M}=0.3153$ as favoured by the Planck observations \citep{Planck:2018vyg}. The quantity $z_{\star}$ is the pivot redshift set to $0.5$ in this work\footnote{The choice of $z_{\star}$ can be arbitrary in principle, and it has almost no impact on the final reconstruction result, as demonstrated in \cite{GG24}.}. The Hubble expansion rate $H$ can be obtained using the relation between $H$ and $D_{\A}$, namely,
\begin{equation}
    \frac{H_{\rm{fid}}(z)}{H(z)}=\alpha_{0}\left[1+\alpha_{1}+(2\alpha_{1}+\alpha_{2})x
    +\left(\frac{3}{2}\alpha_{2}+\frac{1}{2}\alpha_{3}\right)x^{2}
    +\left(\frac{2}{3}\alpha_{3}+\frac{1}{6}\alpha_{4}\right)x^{3}
    +\frac{5}{24}\alpha_{4}x^{4}\right].\label{eq:para-H}
\end{equation} The free parameters $\alpha_i$ can be determined by fitting Eqs. (\ref{eq:para-DA}) and (\ref{eq:para-H}) to distance measurements, generally including BAO and supernovae distances, the observational Hubble data (OHD) \citep{CC} and so forth. Ultimately we can obtain a dark energy shape function defined as,  
\begin{equation} \label{eq:fNorm}
 S[f(a)] \equiv \frac{f(a)-f(a_{\star})}{f^\prime(a_{\star})}, \ \ f(a)\equiv A H^{2}a^{3} =B\left[Xa^3\right]+C. 
\end{equation}
Throughout the paper, the prime denotes a derivative with respect to the scale factor $a$, and $X$ is the dark energy density normalised to unity today\footnote{Note that when fitting the $\alpha$'s in Eqs. (\ref{eq:para-DA}) and (\ref{eq:para-H}) to distance measurements, the derived $X$ may not be positive-definite. Therefore when deriving quantities related to $f'$ including the pressure function and the $g$ function, we apply a prior of $X>0$. We check the posteriors and find that this has a marginal effect on the final result.}. By design, $S[f(a)]$ extracts the shape of $Xa^3$ from data, thus constants $A, B, C$ are irrelevant\footnote{For BAO measurements, $A=r_{\dd}^2, \ B=A H_0^2 (1-\Omega_{\M}), \ C=A H_0^2 \Omega_{\M}$ where $r_{\dd}$ is the sound horizon at decoupling. But these constants have no impact on $S, \ P/P(a_{\star}), g(a)$ and $q(a)$ by design.}. A few important functions that are directly related to dark energy can be derived from $f$, including the normalised dark energy pressure function, \begin{equation}
\frac{P}{P(a_{\star})}\equiv \frac{a_\star^2}{a^2} \cdot \frac{f^{\prime}(a)} {f^{\prime}(a_\star)}=\frac{w(a)X(a)}{w(a_\star)X(a_\star)},\label{eq:P}
\end{equation} and the dark energy characterisation function,
\begin{equation}
g(a)\equiv -\frac{1}{3}\left(1+a\frac{f^{\prime\prime}(a)}{f^{\prime}(a)}\right)
=w-\frac{a}{3}\frac{w^{\prime}}{w}\label{eq:g}
\end{equation}
Given $g(a)$, a general solution to the differential equation (\ref{eq:g}) is,
\begin{equation}
w(a)=\frac{w(a_{\B}) f_1(a_{\B}, a)}{1-w(a_{\B}) f_2(a_{\B}, a)}, 
\end{equation} with
\begin{equation}
f_1\left(a_1, a_2\right) \equiv {\rm exp}\left[{-3 \int_{a_1}^{a_2} \frac{g(x)}{x} {\dd}x}\right], \ \ 
f_2\left(a_1, a_2\right) \equiv 3 \int_{a_1}^{a_2} \frac{f_1\left(a_1, x\right)}{x} {\dd} x,
\end{equation} and $w(a_{\B})$ is the boundary condition at $a=a_{\B}$.

In addition, the deceleration function of the Universe can also be derived from $f$, namely,
\begin{equation}
q(a)\equiv -\frac{a}{2}\frac{f^{\prime}(a)}{f(a)}+\frac{1}{2}.\label{eq:q}
\end{equation}

The datasets used in this work include the isotropic and anisotropic BAO measurements listed in Table \ref{table:BAO data}\footnote{Note that the BOSS and WiggleZ surveys have a small overlapping region in footprint thus the BAO measurements from these two surveys are not strictly independent. But given the small overlapping area compared to the full footprint of BOSS, we assume that the cross-covariance in the measurements is negligible.}, the observational $H(z)$ data (OHD) measured using the ages of galaxies \citep{CC}, compiled in \cite{Yu:2017iju}, and a local measurement of $H_0 = 73.04 \pm 1.04 \ {\rm km} \ {\rm s}^{-1} {\rm Mpc}^{-1}$ \citep{Riess:2021jrx}.

\begin{table}
\begin{center}
\begin{tabular}{c|c|c|c|c|c}
\hline\hline
Survey  & $z_{\rm eff}$ & $D_{\rm V}/r_{\rm d}$ & $D_{\rm A}/r_{\rm d}$ & $D_{\rm H}/r_{\rm d}$ & Reference\\
\hline
SDSS  MGS      & $0.15$ & $4.47\pm 0.17$         &{}               &{}      & \cite{Ross:2014qpa}        \\ 
BOSS          & $0.38$ & {}                      & $10.23\pm 0.17$ & $25.00\pm 0.76$ & \cite{BOSS:2016wmc}\\
BOSS          & $0.51$ & {}                      & $13.36\pm 0.21$ & $22.33\pm 0.58$ & \cite{BOSS:2016wmc}\\
eBOSS  LRG        & $0.70$ & {}                      & $17.86\pm 0.33$ & $19.33\pm 0.53$ &  \cite{Bautista:2020ahg}\\
eBOSS ELG        & $0.85$ & $18.33^{+0.57}_{-0.62}$ &{}               &{}             &\cite{deMattia:2020fkb} \\
eBOSS Quasar     & $1.48$ & {}                      & $30.69\pm 0.80$ & $13.26\pm 0.55$  & \cite{Neveux:2020voa}\\
eBOSS Ly$\alpha$        & $2.33$ & {}                      & $37.6\pm 1.9$   & $8.93\pm 0.28$ &  \cite{duMasdesBourboux:2020pck}\\
6dFGS              & $0.106$ & $2.976\pm 0.133$        &{}               &{}          & \cite{2011MNRAS.416.3017B}  \\
WiggleZ             & $0.44$ & $11.495\pm 0.556$       &{}               &{}          & \cite{Kazin:2014qga}  \\
WiggleZ            & $0.60$ & $14.878\pm 0.677$       &{}               &{}          & \cite{Kazin:2014qga}  \\
WiggleZ              & $0.73$ & $16.854\pm 0.576$       &{}               &{}           & \cite{Kazin:2014qga}  \\
DES Y3           & $0.835$& {}                      &{}$18.92\pm 0.51$ &{}            & \cite{DES:2021esc} \\
\hline\hline
\end{tabular}
\caption{A list of BAO datasets used in this work.}
\label{table:BAO data}
\end{center}
\end{table}

We use the {\tt Cobaya} \citep{Torrado_2021} code to sample the parameter space of the $\alpha$'s with the Monte Carlo Markov Chains (MCMC) algorithm, and perform the post-processing using the {\tt Getdist} \citep{Getdist} software.

\section{RESULTS}
\label{sec:result}

\begin{figure}
\includegraphics[width=0.7\textwidth]{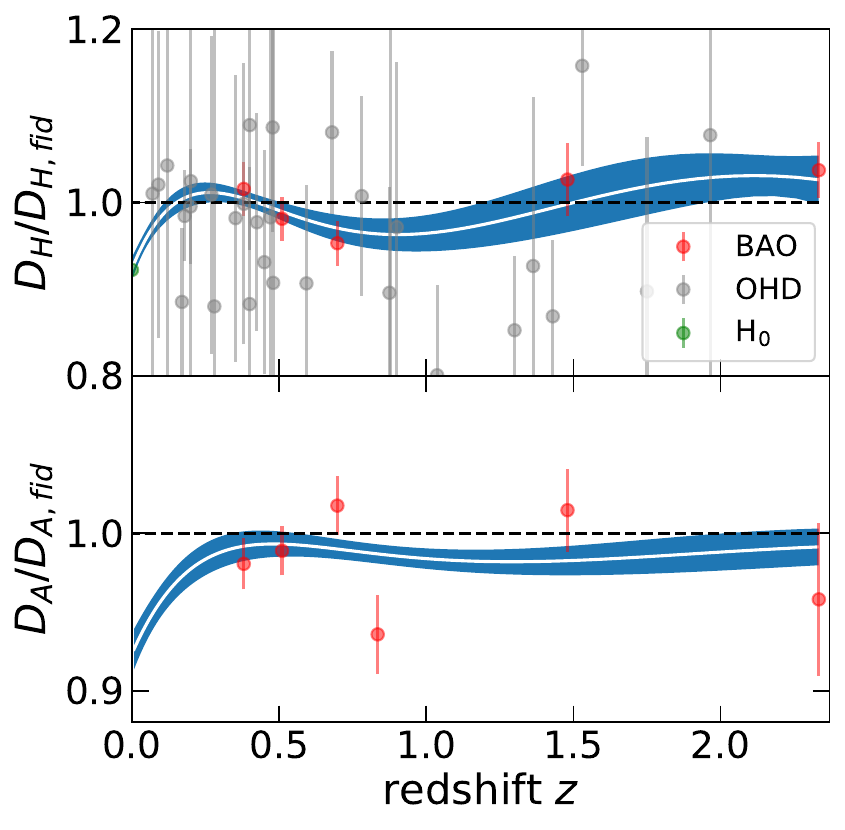}
\centering
\caption{The mean (while curves) and 68\% CL uncertainties of the reconstructed $D_{\HH}/D_{\HH,\fid}$ and $D_{\A}/D_{\A, \fid}$ using the combined dataset of BAO, OHD and $H_0$ measurements shown in data points. The horizontal dashed lines show $D_{\HH}/D_{\HH,\fid}= D_{\A}/D_{\A, \fid}=1$ for a reference.
\label{fig:DH_DM}}
\end{figure}

\begin{figure}
\includegraphics[width=1\linewidth]{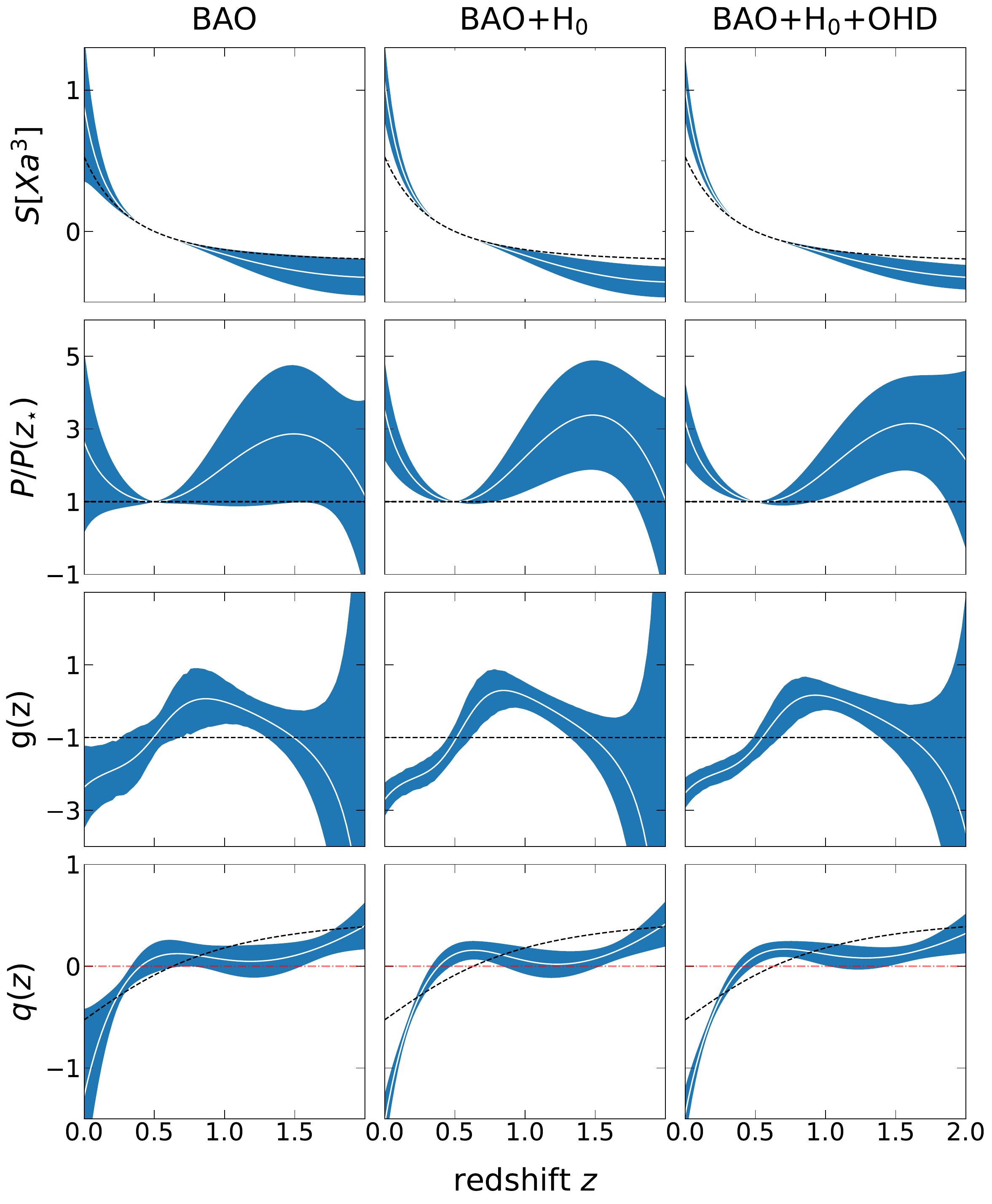}
\centering
\caption{The mean (white curves) and 68\% CL uncertainties (blue shaded regions) of the reconstructed dark energy shape function $S[Xa^{3}]$ (top panels); the normalised dark energy pressure function $P/P(a_{\star})$ (panels in the second rows); the characterisation dark energy function $g(z)$ (third rows), and the deceleration function $q(z)$ of the Universe (bottom panels), derived from three different combinations of datasets illustrated in the legend. In all panels, the black dashed lines show the $\Lambda$CDM prediction, and the red dashed lines in the bottom panels show $q=0$ for a reference. The value of $\Omega_{\M}$ is set to be $0.3153$ when producing the $\Lambda$CDM prediction for $q(z)$.}
\label{fig:4x3}
\end{figure}

\begin{figure}
\includegraphics[width=1.0\textwidth]{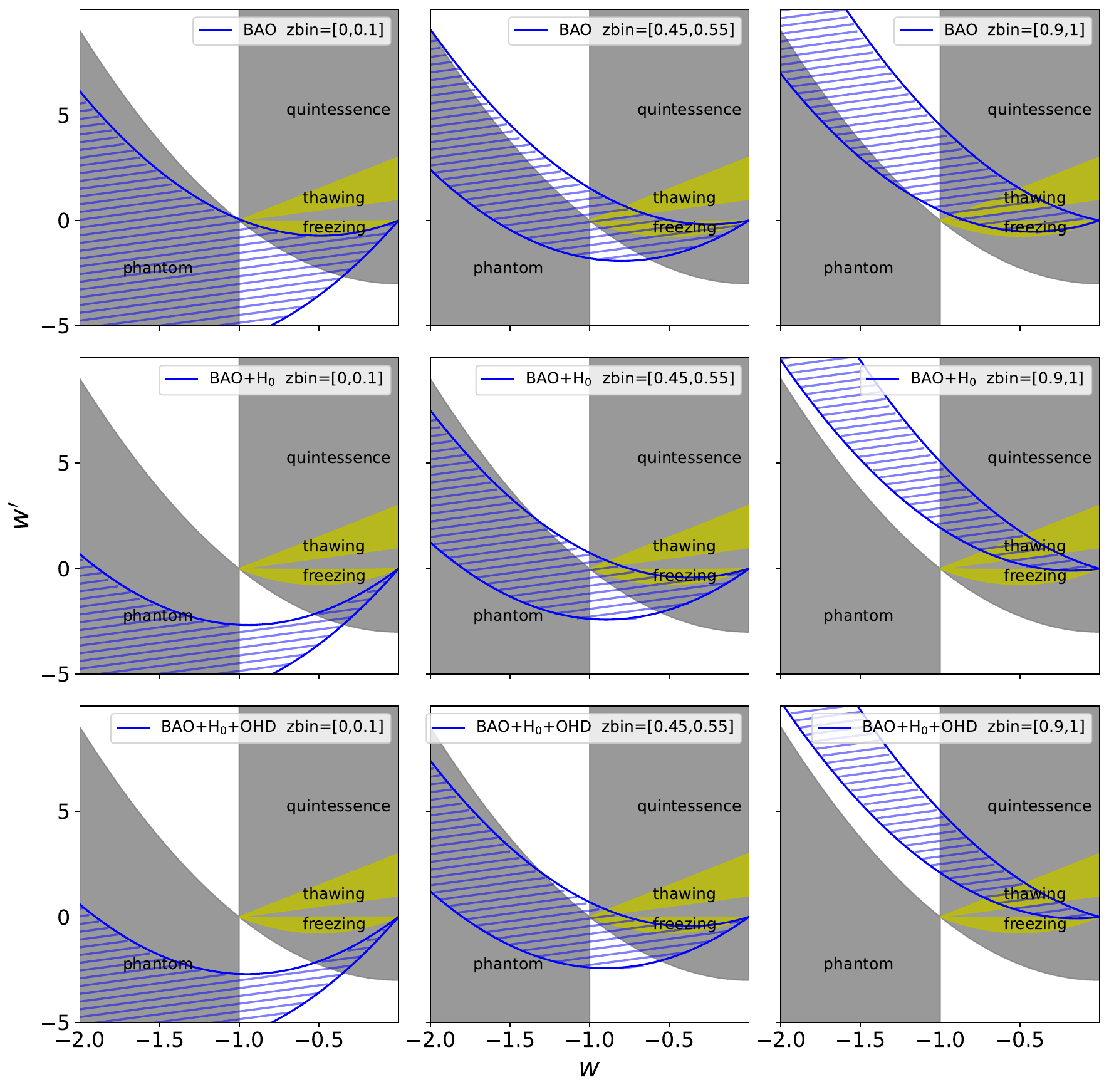}
\centering
\caption{The $w-w'$ phase space diagram predicted by theory models (filled regions) and from observational constraints (hatched regions) through the reconstructed $g(z)$ function derived from three data combinations in three redshift intervals, as shown in the legend.}
\label{fig:wwprime}
\end{figure}

\begin{figure}
\includegraphics[width=\textwidth]{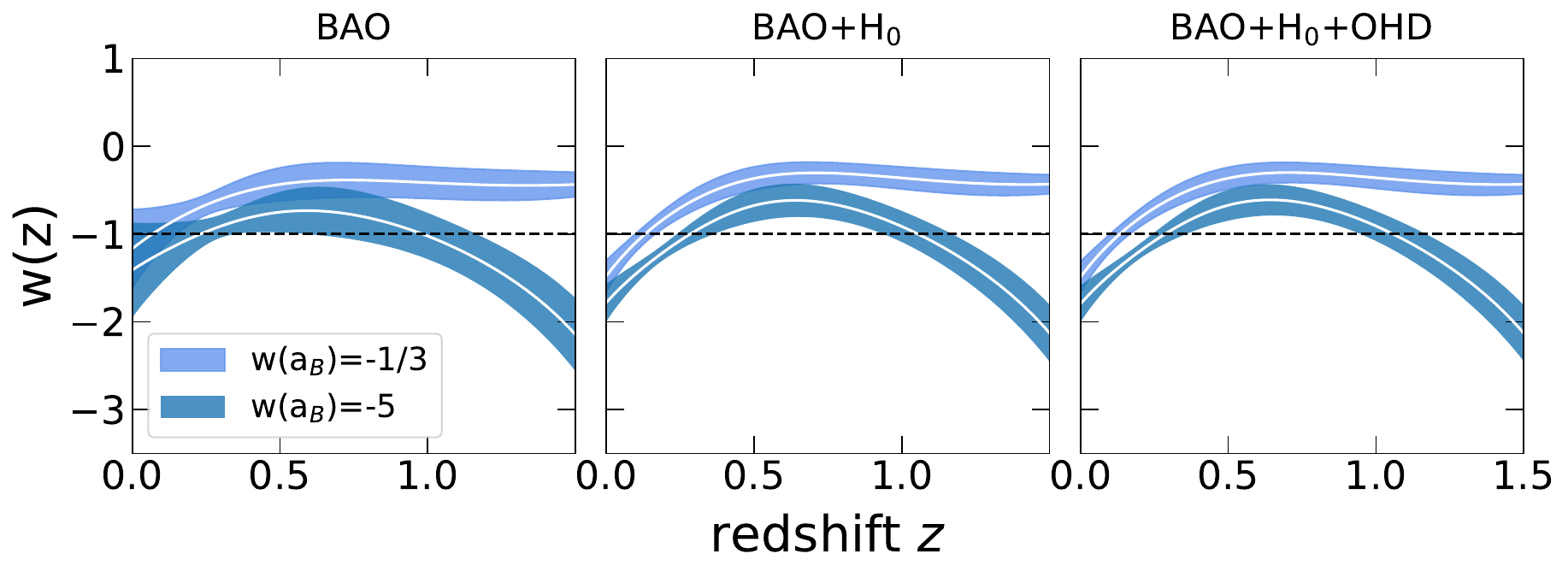}
\centering
\caption{The mean and 68\% CL reconstructed $w(z)$ by solving the differential equation using boundary conditions of $w(a_{\rm B})=-1/3$ and $w(a_{\rm B})=-5$, respectively, where $a_{\rm B}=0.35$. Results shown are derived from different data combinations illustrated in the legend.}
\label{fig:wrecon}
\end{figure}

Our main results are summerised in Figs. \ref{fig:DH_DM} - \ref{fig:wrecon}.

Fig. \ref{fig:DH_DM} shows the mean and 68\% confidence level (CL) reconstructed $D_{\HH}\equiv c/H(z)$ ($c$ is the speed of light) and $D_{\A}$ from the combined dataset of BAO, OHD and $H_0$ measurements, rescaled by the prediction of the fiducial model. As shown, the reconstructed functions significantly deviate from the fiducial $\Lambda$CDM model, and appear wiggly, which is primarily driven by both the $H_0$ and BAO measurements at low and intermediate redshifts, respectively. 


Fig. \ref{fig:4x3} turns the distance measurements into important functions that are directly related to physical features of dark energy. Panels in the top row show the reconstructed dark energy shape function, visualising the time evolution of $Xa^3$. Comparing to the $\Lambda$CDM prediction shown in the black dashed lines, we see that the BAO data tilt $S$ around the pivot point of $z_{\p}=0.5$, namely, the BAO data tend to lift $S$ up at $z<z_{\p}$ and push $S$ downwards at $z>z_{\p}$. Interestingly, this feature is supported by both the $H_0$ and OHD measurements, namely, when $H_0$ and OHD are added to the analysis, the feature becomes more pronounced with much smaller uncertainties, which disfavours the $\Lambda$CDM model.

The $P/P(z_{\star})$ panels provide more information of dark energy. Specifically, we see that $P$ has an apparent local minimum at around $z=0.5$ when the $H_0$ and OHD measurements are combined with the BAO data. This actually disfavours the $w$CDM model, in which $P$ is a power-law function of $a$. 

Further information of $w(z)$ can be learned from the reconstructed $g(z)$ function shown in panels in the third row of Fig. \ref{fig:4x3}. This function shows a strong dynamical behaviour in all cases, indicating an evolving $w(z)$. It is true that solving for $w(z)$ from $g(z)$ requires a boundary condition which is unknown, but we can still use the reconstructed $g(z)$ to identify the allowed region in the $w-w'$ phase space, which can in principle be used to differentiate dark energy models. The shaded area in Fig. \ref{fig:wwprime} illustrate the allowed parameter space by single scalar field models. Specifically, the quintessence \citep{quintessence} and phantom \citep{phantom} models sit in regions of $w>-1 \cap w'>-3(1-w)(1+w)$ and $w<-1 \cap w'<-3(1-w)(1+w)$, respectively \citep{Chiba:2005tj}. The behaviour of the quintessence model may be classified into two scenarios, namely, the thawing ($w\approx-1$ at high redshifts and grows later on) and freezing ($w>-1$ at high redshifts and decreases with time) cases, occupying the regions of $(1+w)<w'<3(1+w)$ and $3w(1+w)<w'<0.2w(1+w)$, respectively \citep{Caldwell:2005tm}. On the other hand, the reconstructed $g(z)$ function constrains the relation between $w$ and $w'$ in a given redshift range, thus we can use $g(z)$ to target the region in the phase space favoured by observations. In Fig. \ref{fig:4x3}, the hatched regions are favoured by observations (at 68\% CL) derived from different data combinations in three redshift intervals, namely, $z\in[0,0.1], \ [0.45,0.55]$ and $[0.9,1.0]$. For all data combinations, we see a similar trend, namely, data favour the phantom model at $z<0.1$, while at $0.9<z<1.0$, most of the data-favoured region overlaps with the quintessence regime. This indicates an evolving $w(z)$ that crosses the $-1$ boundary during evolution, which is consistent with the prediction of the quintom model \citep{quintom}. We check this conclusion by directly solving the differential equation with two choices of boundary conditions, namely, $w(a_{\rm B})=-1/3$ and $w(a_{\rm B})=-5$ with $a_{\rm B}=0.35$, which are sufficiently extreme. The two solutions are shown in Fig. \ref{fig:wrecon}. In all cases, $w(z)$ tends to cross $-1$ during evolution, and the trend becomes significant when $H_0$ and OHD datasets are combined with the BAO data in the analysis.  

Panels in the bottom row of Fig. \ref{fig:4x3} show the reconstructed deceleration function $q(z)$. Comparing to the $\Lambda$CDM prediction with $\Omega_{\M}=0.3153$, we see that data favour a smaller $z_{\rm tr}$, which is the redshift for the acceleration-deceleration transition, namely, $z_{\rm tr}\approx0.4$ while $z_{\rm tr}(\Lambda {\rm CDM})\approx0.6$. This means that the Universe starts accelerating much later than expected in the $\Lambda$CDM model.

\section{CONCLUSION}
\label{sec:conclusion}

Revealing the nature of dark energy is one of the most challenging tasks in modern physics. Equipped with high-quality observational data in this era of precision cosmology, we can use proper theoretical and numerical tools to extract crucial information of dark energy from observations, which can in principle offer important guidance for probing the new physics of dark energy.

In this work, we extract the dark energy shape function, pressure function, the characterisation function and the cosmic deceleration function from measurements of cosmic distances in recent years. Combining the observations of BAO, OHD and $H_0$, we find that $\Lambda$CDM and $w$CDM models are disfavoured, and a dynamical dark energy model with $w$ crossing $-1$ is better supported by data. Interestingly, we find that the cosmic acceleration may have started much later than expected in the $\Lambda$CDM model.

Our method and pipeline is directly applicable to forthcoming distance measurements including BAO measurements from galaxy surveys such as DESI \citep{DESI:2016fyo,DESI:2024mwx}, PFS \cite{PFS} and Euclid \citep{EUCLID}, supernovae surveys of LSST \citep{LSST} and Roman \citep{Roman} and so forth. These planned studies will further investigate the dynamical nature of $w$, which can deepen our understanding of dark energy. 

\newpage

\normalem
\begin{acknowledgements}
We thank Ruiyang Zhao for helpful discussions. All authors are supported by the National Key R \& D Program of China (2023YFA1607800, 2023YFA1607803), NSFC grants (No. 11925303 and 11890691), and by a CAS Project for Young Scientists in Basic Research (No. YSBR-092). SY is also supported by a NSFC grant (No. 12203062). SY and GBZ are also supported by science research grants from the China Manned Space Project with No. CMS-CSST-2021-B01. GBZ is also supported by the New Cornerstone Science Foundation through the XPLORER prize.
\end{acknowledgements}
  
\bibliographystyle{raa}
\bibliography{bibtex}

\begin{thebibliography}{37}
\providecommand\natexlab[1]{#1}
\providecommand\JournalTitle[1]{#1}

\bibitem[Abbott {et~al.}(2022)]{DES:2021esc}
Abbott, T. M.~C., {et~al.} 2022, Phys. Rev. D, 105, 043512

\bibitem[Adame {et~al.}(2024)]{DESI:2024mwx}
Adame, A.~G., {et~al.} 2024, arXiv:2404.03002

\bibitem[Aghamousa {et~al.}(2016)]{DESI:2016fyo}
Aghamousa, A., {et~al.} 2016, arXiv:1611.00036

\bibitem[Aghanim {et~al.}(2020)]{Planck:2018vyg}
Aghanim, N., {et~al.} 2020, Astron. Astrophys., 641, A6, [Erratum:
  Astron.Astrophys. 652, C4 (2021)]

\bibitem[Alam {et~al.}(2017)]{BOSS:2016wmc}
Alam, S., {et~al.} 2017, Mon. Not. Roy. Astron. Soc., 470, 2617

\bibitem[Bautista {et~al.}(2020)]{Bautista:2020ahg}
Bautista, J.~E., {et~al.} 2020, Mon. Not. Roy. Astron. Soc., 500, 736

\bibitem[{Beutler} {et~al.}(2011)]{2011MNRAS.416.3017B}
{Beutler}, F., {Blake}, C., {Colless}, M., {et~al.} 2011, \mnras, 416, 3017

\bibitem[Caldwell(2002)]{phantom}
Caldwell, R.~R. 2002, Phys. Lett. B, 545, 23

\bibitem[Caldwell \& Linder(2005)]{Caldwell:2005tm}
Caldwell, R.~R., \& Linder, E.~V. 2005, Phys. Rev. Lett., 95, 141301

\bibitem[Chiba(2006)]{Chiba:2005tj}
Chiba, T. 2006, Phys. Rev. D, 73, 063501, [Erratum: Phys.Rev.D 80, 129901
  (2009)]

\bibitem[Clifton {et~al.}(2012)]{MG}
Clifton, T., Ferreira, P.~G., Padilla, A., \& Skordis, C. 2012, Phys. Rept.,
  513, 1

\bibitem[{Cole} {et~al.}(2005)]{BAO05b}
{Cole}, S., {Percival}, W.~J., {Peacock}, J.~A., {et~al.} 2005, Mon. Not. Roy.
  Astron. Soc., 362, 505

\bibitem[Copeland {et~al.}(2006)]{DE}
Copeland, E.~J., Sami, M., \& Tsujikawa, S. 2006, Int. J. Mod. Phys. D, 15,
  1753

\bibitem[de~Mattia {et~al.}(2021)]{deMattia:2020fkb}
de~Mattia, A., {et~al.} 2021, Mon. Not. Roy. Astron. Soc., 501, 5616

\bibitem[Di~Valentino {et~al.}(2021)]{Hubble_tension}
Di~Valentino, E., Mena, O., Pan, S., {et~al.} 2021, Class. Quant. Grav., 38,
  153001

\bibitem[du~Mas~des Bourboux {et~al.}(2020)]{duMasdesBourboux:2020pck}
du~Mas~des Bourboux, H., {et~al.} 2020, Astrophys. J., 901, 153

\bibitem[{Eisenstein} \& {Hu}(1998)]{BAO98}
{Eisenstein}, D.~J., \& {Hu}, W. 1998, Astrophys. J., 496, 605

\bibitem[{Eisenstein} {et~al.}(2005)]{BAO05a}
{Eisenstein}, D.~J., {Zehavi}, I., {Hogg}, D.~W., {et~al.} 2005, Astrophys. J.,
  633, 560

\bibitem[Ellis {et~al.}(2014)]{PFS}
Ellis, R., {et~al.} 2014, Publ. Astron. Soc. Jap., 66, R1

\bibitem[Feng {et~al.}(2005)]{quintom}
Feng, B., Wang, X.-L., \& Zhang, X.-M. 2005, Phys. Lett. B, 607, 35

\bibitem[Gu {et~al.}(2024)]{GG24}
Gu, G., Wang, X., Mu, X., Yuan, S., \& Zhao, G.-B. 2024, arXiv:2404.06303

\bibitem[Kazin {et~al.}(2014)]{Kazin:2014qga}
Kazin, E.~A., {et~al.} 2014, Mon. Not. Roy. Astron. Soc., 441, 3524

\bibitem[Laureijs {et~al.}(2011)]{EUCLID}
Laureijs, R., {et~al.} 2011, arXiv:1110.3193

\bibitem[Lewis(2019)]{Getdist}
Lewis, A. 2019, arXiv:1910.13970

\bibitem[{LSST Science Collaboration} {et~al.}(2009)]{LSST}
{LSST Science Collaboration}, {Abell}, P.~A., {Allison}, J., {et~al.} 2009,
  arXiv e-prints, arXiv:0912.0201

\bibitem[Neveux {et~al.}(2020)]{Neveux:2020voa}
Neveux, R., {et~al.} 2020, Mon. Not. Roy. Astron. Soc., 499, 210

\bibitem[{Perlmutter} {et~al.}(1999)]{1999ApJ...517..565P}
{Perlmutter}, S., {Aldering}, G., {Goldhaber}, G., {et~al.} 1999, \apj, 517,
  565

\bibitem[Ratra \& Peebles(1988)]{quintessence}
Ratra, B., \& Peebles, P. J.~E. 1988, Phys. Rev. D, 37, 3406

\bibitem[{Riess} {et~al.}(1998)]{1998AJ....116.1009R}
{Riess}, A.~G., {Filippenko}, A.~V., {Challis}, P., {et~al.} 1998, \aj, 116,
  1009

\bibitem[Riess {et~al.}(2022)]{Riess:2021jrx}
Riess, A.~G., {et~al.} 2022, Astrophys. J. Lett., 934, L7

\bibitem[Ross {et~al.}(2015)]{Ross:2014qpa}
Ross, A.~J., Samushia, L., Howlett, C., {et~al.} 2015, Mon. Not. Roy. Astron.
  Soc., 449, 835

\bibitem[{Spergel} {et~al.}(2015)]{Roman}
{Spergel}, D., {Gehrels}, N., {Baltay}, C., {et~al.} 2015, arXiv e-prints,
  arXiv:1503.03757

\bibitem[Stern {et~al.}(2010)]{CC}
Stern, D., Jimenez, R., Verde, L., Kamionkowski, M., \& Stanford, S.~A. 2010,
  JCAP, 02, 008

\bibitem[Torrado \& Lewis(2021)]{Torrado_2021}
Torrado, J., \& Lewis, A. 2021, Journal of Cosmology and Astroparticle Physics,
  2021, 057

\bibitem[Weinberg(1989)]{Weinberg:1988cp}
Weinberg, S. 1989, Rev. Mod. Phys., 61, 1

\bibitem[Yu {et~al.}(2018)]{Yu:2017iju}
Yu, H., Ratra, B., \& Wang, F.-Y. 2018, Astrophys. J., 856, 3

\bibitem[Zhu {et~al.}(2015)]{Zhu:2014ica}
Zhu, F., Padmanabhan, N., \& White, M. 2015, Mon. Not. Roy. Astron. Soc., 451,
  236

\end{thebibliography}

\end{document}